# Tidal Excitation of Modes in Binary Systems with Applications to Binary Pulsars

Pawan Kumar†, Chi On Ao and Eliot J. Quataert

Department of Physics

Massachusetts Institute of Technology, Cambridge, MA 02139

† Alfred P. Sloan Fellow & NSF Young Investigator

email addresses: pk@brmha.mit.edu (P. Kumar)

coao@alioth.mit.edu (C.O. Ao)

ejquat@brmha.mit.edu (E.J. Quataert)

# ABSTRACT

We consider the tidal excitation of modes in a binary system of arbitrary eccentricity. For a circular orbit, the modes generally undergo forced oscillation with a period equal to the orbital period ($T$). For an eccentric orbit, the amplitude of each tidally excited mode can be written approximately as the sum of an oscillatory term that varies sinusoidally with the mode frequency and a 'static' term that follows the time dependence of the tidal forcing function. The oscillatory term falls off exponentially with increasing $b_\alpha$ (defined as the ratio of the periastron passage time to the mode period), whereas the 'static' term is independent of $b_\alpha$. For small $b_\alpha$ modes ($b_\alpha \approx 1$), the two terms are comparable, and the magnitude of the mode amplitude is nearly constant over the orbit. For large $b_\alpha$ modes ($b_\alpha \gtrsim$ a few), the oscillatory term is very small compared to the 'static' term, in which case the mode amplitude, like the tidal force, varies as the distance cubed. For main sequence stars, $p$, $f$, and low order $g$-modes generally have large $b_\alpha$ and hence small amplitudes of oscillation. High overtone $g$-modes, however, have small overlap with the tidal forcing function. Thus, we expect an intermediate overtone $g$-mode with $b_\alpha \sim 1$ to have the largest oscillation amplitude. In addition, we find that the mode amplitude is independent of the dissipation rate except when the mode frequency is very close to orbital resonance or the damping time is less than $T$; both conditions are unlikely. Moreover, orbital evolution causes a resonant mode to move off resonance with time. This severely limits the amplitude of modes near resonance. Rotation of the star shifts the mode freqeunices but has otherwise little effect on the mode amplitude (provided that the rotation rate is small). Hence, tidally excited modes have amplitudes and phases that are periodic with period $T$, making them readily distinguishable from oscillations excited by other mechanisms.

We apply our work to the SMC radio pulsar PSR J0045-7319, which is believed to be in highly eccentric orbit with a 10 $M_\odot$ B-star. We find that the $g_7$ mode (with period of 1.07 days) of the B-star has the largest oscillation amplitude, with a flux variation of 2.3 m-mag and a surface velocity of 70 m s$^{-1}$. The flux variation at periastron, summed over all modes, is about 10 m-mag; in addition, we propose that the *shape* of the light curve can be utilized to determine the orbital inclination angle. The apsidal motion of this system, calculated without the usual static approximation, is larger than that predicted by the classical apsidal formula by about 1.0%. For the PSR B1259-63 system, the tidal amplitude of the Be-star companion is smaller by a factor of 70 due to its larger periastron distance. To understand the dependence of tidal excitation on stellar structure, detailed numerical calculations of modes of a general polytropic star are also presented.

## §1. Introduction

The study of tidal interactions in binary systems is, among other things, useful as a test of stellar structure (the apsidal motion test, cf. Schwarzschild 1958; Claret and Giménez 1993), as a mechanism for the formation of close binaries in globular clusters (Fabian et al. 1975; Press and Teukolsky 1977; Lee and Ostriker 1986; Kochanek 1992), and for explaining the synchronization and circularization of orbits (Lecar et al. 1976; Zahn 1977, 1989; Hut 1981). The primary focus of this paper, however, is the study of the tidally induced oscillations of a star and their use as a possible probe of stellar or orbital parameters.

Our work is in part motivated by the recent discovery of two radio pulsars in binary systems with main sequence stars (PSR J0045-7319, see Kaspi et al. 1994; and PSR B1259-63, see Johnston et al. 1994). In the past, radio pulsar timing observations have been used to determine orbital parameters with unprecedented precision (Taylor and Weisberg 1989; Kaspi et al. 1994) Since both stars in the binary systems were compact, tidal interactions could be neglected. Indeed, the absence of tidal effects in binary pulsar systems has allowed various general relativistic effects to be measured. The discovery of binary radio pulsars with normal star companions, however, has opened up the possibility for measuring tidal interactions with great precision to probe the interior of normal stars. PSR J0045-7319, because of its high eccentricity and small periastron distance, is currently the best candidate for finding tidally excited oscillations. In fact, PSR J0045-7319 has one of the smallest values for the periastron distance (in units of stellar radius) of all the known detached binary systems.

Tidal deformation of a star can be modeled as a linear superposition of its non-radial pressure and gravity normal mode oscillations (Press and Teukolsky 1977). We investigate how individual mode amplitudes vary with modal and orbital parameters for a Keplerian orbit with arbitrary eccentricity. The dependence of the mode amplitude on the rotation of the star will also be discussed. We ignore, however, the new family of modes that appears in a rotating star (see Papaloizou and Pringle 1978, for a discussion of the rotation modes).

We calculate the surface oscillation amplitude of the star as well as the perturbation to the pulsar orbit as a possible means for detecting these oscillations. We also calculate the detailed shape of the light curve near periastron, which can be used to determine the orbital inclination angle for eccentric binaries such as PSR J0045-7319.

The plan of this paper is as follows. In the next section we establish the main equation for the tidal excitation of modes and discuss, in general, the energy in tidally excited modes, the phenomena of tidal resonance, and the effects of rotation and orbital evolution on the mode amplitude. In §3 we discuss the relation between the mode amplitude and the more observationally accessible quantities such as flux variation, surface velocity, and perturbation to the pulsar's orbit. In §4 we show how the light curve at periastron can be used to determine the orbital inclination angle. In §5 we apply our work to two binary radio pulsars and to general polytropic stars. Our results are summarized in §6.

## §2 Formalism

In this section we present our basic method for studying the tidal excitation of modes in rotating and non-rotating stars in orbits of arbitrary eccentricity, and discuss the significance of the various modal and orbital parameters. We further present order of magnitude estimates for the mode energy.

### §2.1 Basic equation for tidal excitation & its general solution

Consider a binary system where the primary star, taken to be a point mass, has mass $M_p$ and its companion star, the secondary, has mass $M_*$. The position of the primary star in the orbit is specified by $R(t)$, the separation between the stars, and the azimuthal angle, $\phi(t)$. The secondary is assumed to be rotating as a solid body with angular speed $\Omega_*$ about an axis perpendicular to the orbital plane.

The perturbation to any physical quantity in the star, for instance the Eulerian density perturbation, due to the tidal gravitational field of the primary, can be written as a sum over its normal mode amplitudes

$$\delta\rho(\mathbf{r}) = \sum_{n\ell m} A_{n\ell m} \delta\rho_{n\ell}(r) Y_{\ell m}(\theta, \phi), \tag{1}$$

where $\delta\rho_{n\ell}$ is the normalized density eigenfunction (normalized such that the energy in the mode is unity), $Y_{\ell m}(\theta, \phi)$ is the spherical harmonic function, and $A_{n\ell m}$ is the mode amplitude. The pressure ($p$) and internal gravity ($g$) modes are uniquely specified by three numbers: $n$, the number of nodes in the radial direction, the spherical harmonic degree $\ell$, and the azimuthal number $m$; the surface gravity mode, or the fundamental mode ($f$), has no nodes along the radial direction. We use a shorthand notation, $\alpha$, to denote the collective index $(n, \ell, m)$.

The equation for the mode amplitude, $A_\alpha$, can be cast in the form of the following second order ordinary differential equation for a forced, damped harmonic oscillator:

$$\frac{d^2 A_\alpha}{dt^2} + 2\Gamma_\alpha \frac{dA_\alpha}{dt} + \omega_\alpha^2 A_\alpha = f_\alpha(t), \tag{2}$$

with

$$f_\alpha(t) = \frac{4\pi}{(2\ell+1)} \frac{GM_p \omega_\alpha^2 Q_{nl}}{R^{\ell+1}(t)} Y_{\ell m}^*(\theta_{orb} = \pi/2, \phi_{orb}(t) - \Omega_* t), \tag{3}$$

as derived for a non-rotating star in Press and Teukolsky (1977), where

$$Q_{nl} \equiv \int_0^{R_*} dr\, r^{\ell+2} \delta\rho_\alpha(r) \tag{4}$$

is the overlap integral for the mode, $\Gamma_\alpha$ is its dissipation rate, $R_*$ is the radius of the star, and the orbital plane is taken to be at $\theta_{orb} = \pi/2$. It should be noted that our $Q_{n\ell}$ is equal to Press and Teukolsky's overlap integral divided by $\omega_\alpha$. This is due to the difference in the normalization of our eigenfunctions. Our normalized modes have unit energy, whereas Press and Teukolsky's have energy equal to $\omega_\alpha^2$.

It is useful to define the following reduced forcing function

$$\tilde{f}_{\ell m}(t) = \frac{Y_{\ell m}^*(\pi/2, \phi_{orb}(t) - \Omega_* t)}{[R(t)/a]^{\ell+1}}, \tag{5}$$

where $a$ is the semi-major axis of the orbit. The function $\tilde{f}_{\ell m}(t)$ contains all the time dependence of $f_\alpha(t)$ and is independent of the mode properties. Since the driving term falls off as $1/R^{\ell+1}$, most of the energy is in the lowest degree modes that can be tidally excited, i.e., the quadrupole modes. We rewrite $f_\alpha(t)$ in the following compact form:

$$f_\alpha(t) = f_{cons} K_\alpha \Omega_o^2 \tilde{f}_{\ell m}, \tag{6}$$

where

$$f_{cons} \equiv \left[\frac{4\pi}{(2\ell+1)a^{\ell-2}}\right]\left(\frac{M_p}{M_t}\right), \qquad K_\alpha = \omega_\alpha^2 Q_{n\ell}, \tag{7}$$

$\Omega_o = \sqrt{GM_t/a^3}$ is the orbital frequency, and $M_t$ is the total mass of the binary system. All the modal information is contained in the constant factor $K_\alpha$. For quadrupole modes, considered in detail here, $f_{cons}$ is equal to $4\pi/5$ times the fractional mass of the primary star.

Equation (2) can be solved in terms of a Green's function to yield

$$A_\alpha(t) = f_{cons} \left(\frac{K_\alpha}{\omega_\alpha'}\right) \Omega_o^2 \int_{-\infty}^t dt_1\, \exp[-\Gamma_\alpha(t-t_1)]\, \sin[\omega_\alpha'(t-t_1)]\, \tilde{f}_{\ell m}(t_1), \tag{8}$$

where $\omega_\alpha' \equiv \sqrt{\omega_\alpha^2 - \Gamma_\alpha^2} \approx \omega_\alpha$.

While equation (8) allows us to calculate the mode amplitude for a star with arbitrary rotation rate, it does not lend itself to a simple interpretation of the significance of various parameters. However, a simple interpretation is possible when the forcing function is periodic, such as when the star is non-rotating, and one can analyze the mode amplitude using a Fourier series. In the following discussion we assume that the star is non-rotating. The effects of rotation on the mode frequency and amplitude are addressed in §2.3.

When rotation of the star is neglected, the reduced forcing function $\tilde{f}_{\ell m}(t)$ is periodic with orbital period $T$, and we decompose it in terms of its Fourier coefficients:

$$\tilde{f}_{\ell m}(t) = \sum_{n=1}^{\infty} C_n^{(\ell m)} \sin(n\Omega_o t) + \sum_{n=0}^{\infty} D_n^{(\ell m)} \cos(n\Omega_o t). \tag{9}$$

The Fourier coefficients $C_n^{(\ell m)}$ and $D_n^{(\ell m)}$ depend only on the eccentricity, $e$. The mode amplitude can easily be calculated using this series and is given below

$$\frac{A_\alpha(t)}{f_{cons} K_\alpha} \equiv \tilde{A}_\alpha(t) = \sum_{n=1}^{\infty} \frac{C_n^{(\ell m)} \sin(n\Omega_o t + \phi_n)}{\left[(n^2 - r_\alpha^2)^2 + 4n^2 d_\alpha^2\right]^{1/2}} + \sum_{n=0}^{\infty} \frac{D_n^{(\ell m)} \cos(n\Omega_o t + \phi_n)}{\left[(n^2 - r_\alpha^2)^2 + 4n^2 d_\alpha^2\right]^{1/2}}, \tag{10}$$

where

$$\tan \phi_n = -\frac{2n d_\alpha}{(r_\alpha^2 - n^2)}, \tag{11}$$

and the dimensionless numbers $r_\alpha$ and $d_\alpha$ are defined below:

$$r_\alpha \equiv \frac{\omega_\alpha}{\Omega_0}, \qquad d_\alpha \equiv \frac{\Gamma_\alpha}{\Omega_0}. \tag{12}$$

It is clear from equation (10) that resonances occur for integral values of $r_\alpha$. Because of our choice of normalization for the eigenfunctions (to yield unit energy), the energy in the modes, $E_\alpha$, is approximately equal to $|A_\alpha|^2$.

For highly eccentric orbits, the tidal force is appreciable only when the star is near periastron. Therefore, another relevant time scale is the periastron passage time. We define a dimensionless parameter below which expresses the mode frequency in terms of this time scale,

$$b_\alpha \equiv \frac{\omega_\alpha}{\Omega_p}, \tag{13}$$

where $\Omega_p$ is the angular velocity of the star at periastron and is given by

$$\Omega_p = \frac{(1-e^2)^{\frac{1}{2}}}{(1-e)^2} \Omega_o. \tag{14},$$

The amplitude for a quadrupole mode is equal to $(4\pi/5)(M_p/M_t)\tilde{A}_\alpha K_\alpha$ (see eq. [10]). The dependence of $A_\alpha$ on the orbital parameters is entirely contained in $\tilde{A}_\alpha$, which is a function of dimensionless parameters, $e$, $r_\alpha$ and $d_\alpha$; its properties are discussed below. In addition, $A_\alpha$ depends on the structure of the star, via the parameter $K_\alpha$, which will be calculated in §5. The relation of $A_\alpha$ to the observed quantities is non-trivial and is discussed in §3.

The behavior of $\tilde{A}_\alpha$ can be understood in terms of the Fourier coefficients. It is easy to show that for quadrupole tide $D_n$'s are real, $C_n$'s are imaginary, and $C_n^{(20)} = 0$. In addition, numerical calculations show that $|C_n^{(22)}| \sim |D_n^{(22)}|$ for large $n$ ($n \gtrsim \Omega_p/\Omega_o$). Moreover, $D_n^{(20)}$ falls off as approximately $\exp(-\alpha_e n\Omega_o/\Omega_p)/(1-e)^{3/2}$ for $n \gtrsim 5\Omega_p/\Omega_o$, where $\alpha_e \approx 1.3e^{-0.25}$. The Fourier coefficients for $m=\pm 2$ are larger compared to $m=0$ by a factor of about $\exp(2\alpha_e)$. The exponential dependence arises because for eccentric orbits, the characteristic time scale for the variation of the tidal force is the periastron passage time, $\Omega_p^{-1}$; therefore, the power spectrum for frequencies greater than $\Omega_p$ falls off exponentially.

Tidally excited modes of a star in a circular orbit exhibit oscillations with a period equal to that of the orbit and not their natural oscillation periods (so long as the orbital evolution timescale is long compared to the mode dissipation time). In the general case of an elliptical orbit, the time dependence of the mode amplitude consists of two parts: an almost static part that varies on the time scale of the tidal force and a part that oscillates with the mode frequency. To be more specific, we can write the reduced mode amplitude as $\tilde{A}_\alpha(t) = \tilde{A}_\alpha^{(r)}(t) + \tilde{A}_\alpha^{(or)}(t)$, where $\tilde{A}_\alpha^{(r)}(t)$ is equal to the sum of the resonant and near resonant terms ($n \approx r_\alpha$) in the Fourier series and $\tilde{A}_\alpha^{(or)}(t)$ is the sum of the off-resonance terms. The former varies sinusoidally with the mode frequency, while the latter is approximately proportional to the tidal forcing function, which peaks strongly near periastron. Since the Fourier coefficients fall off exponentially, the oscillation amplitude, $\tilde{A}_\alpha^{(r)}(t)$, decreases with $r_\alpha$ as $\exp(-\alpha_e r_\alpha \Omega_o/\Omega_p) = \exp(-\alpha_e b_\alpha)$ for $b_\alpha \gtrsim 5$. On the other hand, because most of the contribution to $\tilde{A}_\alpha^{(or)}$ comes from lower order terms in the Fourier series, $\tilde{A}_\alpha^{(or)}$ is approximately proportional to $1/b_\alpha^2$. It should be pointed out that this separation of the mode amplitude into two parts only makes physical sense for modes with periods comparable to or smaller than the periastron passage time (i.e. $b_\alpha \gtrsim 1$). Modes with $b_\alpha \ll 1$ cannot respond quickly enough to the variation of the tidal force near periastron. Hence, $\tilde{A}_\alpha^{(or)}$ for these modes are not expected to be proportional to the forcing function.

Figure 1 shows the time variation of the reduced mode amplitude ($\tilde{A}_\alpha$) for three different values of $b_\alpha$ for the same orbit. For small $b_\alpha$, the resonant term is comparable to the sum of the off-resonance terms. Therefore, $\tilde{A}_\alpha$ does not vary much from periastron to apstron. In contrast, modes with higher $b_\alpha$ have large peaks near periastron, arising from the static term $\tilde{A}_\alpha^{(or)}(t)$, and much smaller oscillation amplitudes due to the exponential fall-off of $\tilde{A}_\alpha^{(r)}(t)$ with $b_\alpha$. The time variation of the reduced mode amplitude for large $b_\alpha$, being dominated by the static part, follows closely that of the driving function. Physically, this behavior is due to the fact that for large $b_\alpha$, the oscillator adjusts adiabatically to the driving function; hence, the mode amplitude is approximately equal to the driving function divided by the mode frequency squared. Even the $b_\alpha=10$ mode does, however, have a non-zero oscillatory component, but its amplitude is a factor of about 2000 smaller compared to the static tidal amplitude at periastron. Thus it is exceedingly difficult to observe pulsation of modes with large $b_\alpha$ values.

The dependence of $\tilde{A}_\alpha$ on $r_\alpha$ and $d_\alpha$ is simple to understand by making use of equation (10). We consider the various possibilities here. When $r_\alpha$ is a 'small' integer, that is, when $r_\alpha \lesssim 5\Omega_p/\Omega_o$ or $b_\alpha \lesssim 5$, and $d_\alpha \lesssim 1$, the $r_\alpha$-th Fourier coefficient is not much smaller than the first, and the contribution to the mode amplitude is dominated by the resonant term. In this case the oscillatory part of the mode amplitude is much greater than the static part, and the mode energy decreases with increasing damping rate as $1/d_\alpha^2$. For $d_\alpha \gg 1$, several harmonics of the orbital frequency lie within the modal linewidth, and the mode energy is proportional to $1/d_\alpha$.

For 'large' integral values of $r_\alpha$, namely $r_\alpha \gtrsim 5\Omega_p/\Omega_o$ or $b_\alpha \gtrsim 5$, the Fourier coefficients $C_{r_\alpha}^{(\ell m)}$ and $D_{r_\alpha}^{(\ell m)}$ are so small (due to the exponential fall–off) that off–resonance terms in the series make the dominant contribution to $\tilde{A}_\alpha$. In this regime, the static tide limit, the oscillatory part of the amplitude is much smaller than the static amplitude. In addition, the mean energy in the tides is essentially independent of the damping rate.

Finally, for non-integral values of $r_\alpha$, it is convenient to write $r_\alpha$ as the sum of the nearest integer and $\delta r_\alpha$. For $d_\alpha \ll \delta r_\alpha$ the mode amplitude is independent of the damping rate, and increases as $1/\delta r_\alpha$. This has the obvious physical interpretation that the amplitude of an oscillator driven off-resonance is independent of its damping. For $d_\alpha \gg \delta r_\alpha$ the behavior is the same as the case of integer $r_\alpha$ discussed above.

The oscillation amplitude is proportional to the product of $\tilde{A}_\alpha^{(r)}$ and the overlap integral ($Q_\alpha$). There are thus two opposing effects in the tidal excitation of modes. On the one hand, larger $n$ g-modes have smaller frequencies and thus smaller values of $b_\alpha$; they therefore have larger values of $\tilde{A}_\alpha^{(r)}$. On the other hand, $Q_\alpha$ are larger for modes with smaller $n$; this condition favors low order modes. For these competing reasons, the modes with the largest oscillation amplitude, for most binary systems, are some intermediate order g-mode ($n \sim 5$) with $\delta r_\alpha$ close to zero (see §2.2). It should be noted that the amplitude near periastron is dominated by the $p_1$, $f$, and $g_1$ modes because of their large overlap integrals. Since the values of $b_\alpha$ are generally large for these modes, the time dependence of their amplitudes is the same as the tidal force. Thus we cannot infer much about the star from the observation of its light curve near periastron, though it can be utilized to determine the orbital inclination angle (see §4).

To give a feel for the physically relevant range of parameters, we note that the periods of low $n$ modes for main sequence stars are of order a few hours to a few days, and the damping times are of order 100 years or greater. The energy in these modes due to tidal forcing is independent of damping unless their $b_\alpha$ value are of order unity and $r_\alpha$ happens to be extremely close to an integer ($\delta r_\alpha < d_\alpha$).

## §2.2 Order of Magnitude Estimate of Mode Energy

In this section, we estimate mode frequencies and overlap integrals and use them to calculate the energy in tidally excited modes. The frequencies of quadrupolar $f$ and low order $p$-modes are approximately equal to (Christensen-Dalsgaard and Berthomieu 1991)

$$\omega_{n2} \approx 6^{1/4}(n+1)\left(\frac{GM_*}{R_*^3}\right)^{1/2},$$

where $M_*$ and $R_*$ are the mass and the radius of the star, and $n=0$ for the $f$-mode. The $f$ and $p$-mode frequencies have a weak dependence on the structure of the star which is not included in the above expression. Gravity waves can propagate only in convectively stable media and thus the $g$-mode frequencies are sensitive to stellar structure. For a polytropic star of index $n_{pi}$, and a constant ratio of specific heats, $\gamma$, the $g$-mode frequencies are given below (Christensen-Dalsgaard and Berthomieu 1991)

$$\omega_{n2} \approx \frac{f_g}{(n+1)}\left(\frac{GM_*}{R_*^3}\right)^{1/2},$$

where $f_g = 15[n_{pi}(\gamma-1)-1]/(n_{pi}+1)$; the factor of 15 has been obtained by a fit to numerically calculated frequencies.

Since the eigenfunctions are normalized such that there is unit energy in the mode, the overlap integral, $Q_{n2}$, is proportional to $(R_*^5/G)^{1/2}$. For $n \neq 0$ the wave function is oscillatory in the radial direction and this leads to cancellation of the tidal force experienced by the mode. Thus the overlap integral decreases with $n$. We model this dependence crudely as $(n+1)^{-\beta}$ and write $|Q_{n2}| \sim f_q(R_*^5/G)^{1/2}/(n+1)^\beta$. Numerical calculation of $Q_{n2}$ for polytropic stars with $\gamma = 5/3$ gives $f_q \sim 1$ for $f$ and $p$-modes, and $f_q \sim 1/3$ for g-modes. The exponent, $\beta$, for g-modes varies from 1.5 to 3 as the polytropic index decreases from 3 to 2 (for an analytical derivation of these results, see Zahn 1970), and $\beta=6$ for $f$ and $p$ modes of $n_{pi}=1.5$ stars.

We estimate $\tilde{A}_\alpha(t)$ using equation (10) and find that the energy of oscillations in $m=0$ quadrupole modes is

$$E_\alpha = |A_\alpha|^2 \sim 6\left(\frac{M_p^2}{M_t^2}\right)\frac{D_{n_\alpha}^2 Q_\alpha^2 \omega_\alpha^2 \Omega_o^2}{[(\delta r_\alpha)^2 + d_\alpha^2]},$$

where $n_\alpha$ is the integer nearest to $r_\alpha$. Making use of the above expressions for $Q_\alpha$ and $\omega_\alpha$ of quadrupole $g$-modes and the expression for the Fourier coefficients obtained in the previous section, we find the oscillation energy in $m = 0$ $g$-modes to be

$$E_{n20} \sim E_{orb}\left[\frac{M_p}{M_t}\right]\left(\frac{R_*}{d_{peri}}\right)^2\left[\frac{f_g^2}{(n+1)^{2+2\beta}}\right]\frac{\exp(-2.6 b_\alpha e^{-0.25})}{[(\delta r_\alpha)^2 + d_\alpha^2](1-e)}, \qquad (15)$$

where $d_{peri} = a(1-e)$ is the periastron distance, and $E_{orb} = GM_*M_p/2a$ is the orbital energy of the binary system.

Since $p$-modes have larger frequencies, and thus larger values of $b_\alpha$, than $g$-modes, $p$-mode oscillation energies are generally much smaller than $g$-mode energies. Moreover, we see from equation (15) that the oscillation energy in $g$-modes falls off as approximately $(n+1)^{-5}$ for polytropic stars of $n_{pi}=3$. Thus the mode with the largest oscillation energy is typically a $g$-mode of moderate $n$ value. The energy in $m=\pm 2$ modes is larger compared to $m=0$ modes by a factor of approximately $\exp(5\,e^{-0.25})$.

If the periastron passage time is large compared to the periods of low order $p$ and $g$-modes, the tide near periastron is almost static. In this case we can easily estimate the mode amplitude by discarding the time derivative terms in equation (2). We find that the energy in the tide near periastron, which is dominated by $f$ and low order $p$ and $g$ modes, is equal to $k\,E_{orb}(M_p/M_*)(R_*/d_{peri})^5$, where $k \equiv (2\pi G/5R_*^5)\sum_n Q_{n2}^2$ can be shown to be equal to the apsidal motion constant.

Thus far we have discussed the dependence of the mode amplitude and the average energy on the orbital parameters and the mode frequency (through $r_\alpha$, $b_\alpha$ and $e$) as well as simple order of magnitude estimates of the dependence on stellar structure. A more detailed discussion of the dependence on stellar structure will be given in §5.2.

### §2.3 Effect of stellar rotation on the mode amplitude

Rotation of the star has been neglected in the previous sections. We now discuss the effect of rotation on the mode amplitude. Assuming that the rotation rate is small compared to the mode frequency, we can treat its effects using perturbation theory. To linear order in the rotation rate, the frequencies of the modes, as seen in the rotating frame of the star, are shifted due to the coriolis force and are given by (Unno, et al. 1989)

$$\omega_\alpha^r = \omega_\alpha + \delta\omega_\alpha = \omega_\alpha - m\Omega_*\omega_\alpha^2 \int dr\, r^2 \rho \left[2\xi_{n\ell}^r \xi_{n\ell}^h + (\xi_{n\ell}^h)^2\right] \equiv \omega_\alpha - m\Omega_* J_{n\ell}. \qquad (16)$$

where $\xi_{n\ell}^r$ and $\xi_{n\ell}^h$ are the radial and transverse components of the displacement eigenfunction (see eq. [34]). Together with equation (8), it is clear that to first order, stellar rotation has no effect on modes with $m=0$. (For simplicity, we have ignored the perturbation to the eigenfunctions, which could be substantial and would modify the overlap integral.) To find the effect of rotation on the amplitudes of tidally excited modes with $m \neq 0$, we substitute the Fourier series expansion of the reduced forcing function (eq. [9]) into equation (8) and carry out the integration which yields the following expression for the reduced mode amplitude

$$\tilde{A}_\alpha = \frac{\exp(im\Omega_* t)}{2} \left[ \sum_{n=1}^\infty \frac{(D_n^{(\ell m)} + iC_n^{(\ell m)})\exp(-in\Omega_0 t - i\phi_n)}{\sqrt{[r_\alpha^{r\,2} - (n-ms)^2]^2 + 4d_\alpha^2(n-ms)^2}} \right.$$
$$+ \sum_{n=1}^\infty \frac{(D_n^{(\ell m)} - iC_n^{(\ell m)})\exp(in\Omega_0 t - i\phi'_n)}{\sqrt{[r_\alpha^{r\,2} - (n+ms)^2]^2 + 4d_\alpha^2(n+ms)^2}} \quad (17)$$
$$\left. + \frac{2D_0^{(\ell m)}}{r_\alpha^{r\,2} - m^2 s^2 + 2imd_\alpha s} \right]$$

where

$$r_\alpha^r = \frac{\omega_\alpha^r}{\Omega_o}, \quad s = \frac{\Omega_*}{\Omega_o}, \quad \tan\phi_n = \frac{2d_\alpha(ms-n)}{r_\alpha^{r\,2} - (n-ms)^2}, \quad \tan\phi'_n = \frac{2d_\alpha(ms+n)}{r_\alpha^{r\,2} - (n+ms)^2} \quad (18)$$

It is easy to show that for $m = \pm 2$, $Re(C_n^{(\ell m)}) = Im(D_n^{(\ell m)}) = 0$, and for $n$ greater than a few $Im(C_n^{(\ell m)}) \approx -\text{sign}(m) Re(D_n^{(\ell m)})$. Making use of these in equation (17) we obtain

$$\tilde{A}_{\pm|m|} \approx \exp(im\Omega_* t) \sum_{n=0}^\infty \frac{D_n^{(\ell m)} \exp(\mp in\Omega_0 t \mp i\phi_n(|m|))}{\sqrt{[r_\alpha^{r\,2} - (n-|m|s)^2]^2 + 4d_\alpha^2(n-|m|s)^2}}. \quad (19)$$

The mode amplitude observed from an inertial frame is proportional to $\tilde{A}_\alpha(t)\exp(-im\Omega_* t)$. Thus the observed mode amplitude is a periodic function with phase coherence from one orbit to another, in spite of the rotation of the star. From the denominator in equation (19), we note that resonance occurs when $n = r_\alpha^r + |m|s$. Thus, so long as $\Omega_* \ll \omega_\alpha$, the observed mode amplitude is approximately the same as the amplitude of a mode for a nonrotating star with frequency equal to $\omega_\alpha + \Omega_*(|m| - mJ_{n\ell})$.

### §2.4 Effect of Orbital Evolution on Mode Amplitude Near Resonance

Orbital evolution causes a resonant mode to move off resonance with time. Thus amplitudes of resonant modes do not diverge even in the limit of vanishing damping. We showed in §2.1 that the mode amplitude is proportional to $1/\delta r_\alpha$ if $d_\alpha \ll \delta r_\alpha$. Physically this is the result of the mode amplitude building up as the mode is kicked in phase at periastron over $1/\delta r_\alpha$ orbits. If the change in the orbital period is $\Delta T$ in one orbit, then the phase difference of a mode at periastron in two consecutive orbits is $\omega_\alpha \Delta T$. Therefore, the tidal force will be out of phase with the mode in $2\pi/\sqrt{\omega_\alpha \Delta T}$ orbits. This implies that there is an effective lower limit to $\delta r_\alpha$, $\delta r_{min}$, for the purpose of calculating the amplitude of tidally excited modes. A straightforward calculation yields

$$\delta r_{min} \approx \left[\frac{r_\alpha}{2\pi}\left(\frac{\Delta T}{T}\right)\right]^{1/2}. \quad (20)$$

Thus, because of orbital evolution, it is not physically meaningful to consider resonance where the ratio of the orbital period to the mode period is any closer to an integer than the value $\delta r_{min}$. We calculate $\Delta T$ below, which we then substitute into equation (20) to determine $\delta r_{min}$.

The decrease in period $\Delta T$ is related to the decrease in orbital energy in one orbit, $\Delta E_{orb}$, as follows

$$\frac{\Delta T}{T} = -\frac{3}{2}\frac{\Delta E_{orb}}{E_{orb}}. \qquad (21)$$

The change in the orbital energy is equal to the work done on the star by the tidal force, which is

$$\Delta E_{orb} = -\int_0^T dt \int d^3r\, \mathbf{v} \cdot \mathbf{F}_{tide} = 4\pi G M_p \sum_{n,\ell,m} \frac{Q_{n\ell}}{(2\ell+1)} \int_0^T dt\, \frac{Y^*_{\ell m}(\pi/2, \phi_{orb})}{R^{\ell+1}(t)} \frac{dA_{n\ell m}}{dt}, \qquad (22)$$

where $\mathbf{v}$ is the velocity induced in the star by the tidal force and is equal to $\sum_\alpha \boldsymbol{\xi}_\alpha dA_\alpha/dt$. We calculate the time integral in the above equation using equations (9) and (10) and the result, for $b_\alpha \gg 1$ for the low order modes, i.e. the static tide limit, is given below

$$\frac{\Delta E_{orb}}{E_{orb}} = 32\pi k\, h(e) \left[\frac{\Gamma_{tide}}{\Omega_o}\right] \left(\frac{M_p M_t}{M_*^2}\right)\left(\frac{R_*}{a}\right)^8, \qquad (23)$$

where $k$ is the apsidal motion constant given by

$$k = \frac{2\pi G}{5R_*^5}\sum_n Q_{n2}^2, \qquad (24)$$

$\Gamma_{tide}$ is the dissipation rate of the tide which is in general different from the mode dissipation rate introduced earlier and

$$h(e) \equiv \frac{3072 + 47616 e^2 + 97920 e^4 + 35520 e^6 + 1200 e^8}{2048(1-e^2)^{15/2}}. \qquad (25)$$

In deriving equation (23) we took $\Gamma_q/\omega_q^2 = \Gamma_{tide}(R_*^3/GM_*)$. This is a crude description for the dissipation of static tides in terms of the normal modes of the star and corresponds to assuming that the lag-angle of the star associated with different modes is equal. (It is easy to show that the lag-angle associated with a mode, for slowly rotating stars, is $2\Gamma_q\Omega_0/\omega_\alpha^2$). We note that our expression for orbital energy change (eqs. [23]–[25]) is identical to that of Hut (1981) for nonrotating stars.

Combining equations (20), (21) and (23) we find the minimum possible value of $\delta r_\alpha$ to be

$$(\delta r_{min})^2 = 24 r_\alpha\, k\, h(e) \left[\frac{\Gamma_{tide}}{\Omega_o}\right]\left(\frac{M_p M_t}{M_*^2}\right)\left(\frac{R_*}{a}\right)^8. \qquad (26)$$

The lag-angle of the tidal bulge, $\delta_{tide}$, is related to $\Gamma_{tide}$: $\delta_{tide} \approx 2\Gamma_{tide}\Omega_o(R_*^3/GM_*)$. This can be used to express the above limit in terms of the lag-angle.

The angular momentum of the orbit also changes due to tidal interaction, which causes a change in the orbital eccentricity. However, in the absence of energy transfer from the orbit to the star the orbital period remains unchanged. Therefore, a change in $e$ alone does not change the phase relationship of a mode and the forcing function at periastron. Thus we do not expect our result given in equation (26) to be significantly modified when angular momentum transfer is included. Finally, we note that it can be easily shown that the apsidal motion of the system has the same effect on the mode amplitude and resonance as the stellar rotation discussed in §2.3.

## §3 Observation of pulsation

The pulsation of a star can be detected by one of the following three techniques: (1) photometric or flux variations; (2) spectroscopic observation, i.e., by measuring the time dependent surface velocity; or (3) for binary pulsar systems, the pulse arrival delay due to the periodic variation of the neutron star orbit caused by the time dependent quadrupole moment of the oscillating star. The mode amplitude calculated in the last section needs to be transformed to yield these observational quantities. The details of these transformations are described below. It should be noted that one can distinguish tidally induced pulsation from the intrinsic pulsation of the star because the former has a definite phase relation to the orbital motion.

### §3.1 Flux variation associated with pulsation

Associated with the tidally excited modes there are variations of the star's luminosity. The observed flux variation arises from the change in the surface temperature of the star as well as the change in its projected surface area. To first order, we can calculate these two effects separately and add them to obtain the total flux variation.

The flux change at the stellar surface, associated with a mode $(n, \ell, m)$, is given by:

$$\delta F_{n\ell m}(R_*, \theta, \phi) = A_{n\ell m}\delta F_{n\ell}(R_*)Y_{\ell m}(\theta, \phi), \tag{27}$$

where $\delta F_{n\ell}(r)$ is the flux variation eigenfunction. To calculate the observed flux variation, $\delta F_{n\ell m}^{obs}$, we need to integrate the above expression over the hemisphere of the star facing us. Thus the expression for the fractional observed flux variation is

$$\frac{\delta F_{n\ell m}^{obs}}{F^{obs}} = \frac{\int d\Omega\, \mu\, f_{LD}(\mu) Y_{\ell m}(\theta, \phi) \delta F_{n\ell}(R_*)}{\int d\Omega\, \mu\, f_{LD}(\mu) F}, \tag{28}$$

where

$$\mu = \mathbf{n}_s \cdot \mathbf{n}_{obs}, \tag{29}$$

$\mathbf{n}_s$ is a unit vector normal to the stellar surface, $\mathbf{n}_{obs}$ is a unit vector pointing to the observer (from the center of the star), $F$ is the unperturbed flux, and $f_{LD}$ is the limb darkening function which can be written as

$$f_{LD}(\mu) = a_o + a_1\mu + a_2\mu^2, \tag{30}$$

where $a_o \approx 0.3726$, $a_1 \approx 0.6500$, and $a_2 \approx -0.0226$ (see, e.g., Kopal 1959). The expression in equation (28) does not include the variation in the projected area of the star, which is calculated below.

Specializing to $\ell = 2$ modes and making use of the last four equations, we find the observed flux for the $m=0$ modes to be

$$\frac{\delta F^{obs}_{n20}}{F^{obs}} = \frac{f_{cons}a_{LD}}{4}\sqrt{\frac{5}{16\pi}}\left[\frac{K_{n2}\delta F_{n2}(R_*)}{F}\right]\tilde{A}_{n20}\left[3\cos^2 i - 1\right], \tag{31}$$

and the observed flux for the sum of the $m=-2$ and $m=2$ modes to be

$$\frac{\delta F^{obs}_{n22}}{F^{obs}} = \frac{f_{cons}a_{LD}}{2}\sqrt{\frac{15}{32\pi}}\left[\frac{K_{n2}\delta F_{n2}(R_*)}{F}\right]\left[Re(\tilde{A}_{n22})\cos(2\phi_0 - 2\Omega_*t) \right. \\ \left. - Im(\tilde{A}_{n22})\sin(2\phi_0 - 2\Omega_*t)\right]\sin^2 i, \tag{32}$$

where $i$ is the inclination angle of the orbit, and $\phi_0$ is the azimuthal angle of the line-of-sight projected onto the orbital plane and is related to the longitude of periastron ($\omega$) by $\phi_0 = \omega - \pi/2$. The factor $a_{LD}$, which arises from limb darkening, is defined as

$$a_{LD} \equiv \frac{15a_o + 16a_1 + 15a_2}{5(3a_o + 2a_1 + 1.5a_2)}. \tag{33}$$

Note that the observed flux variations for $m=0$ and $m=\pm 2$ modes (eqs. [31] & [32]) have a different dependence on the inclination angle. This can be used to determine the orbital inclination angle (see §4 for details).

The calculation of the flux variation eigenfunction, $\delta F_{n\ell}$, requires solving coupled linear differential equations for nonadiabatic oscillations, which is quite involved. However, for most purposes it is reasonably accurate to use the relation $\delta F_{n2}(R_*)/F \approx 4\Delta T(\tau = 2/3)/T$, where $\Delta T(\tau = 2/3)$ is the Lagrangian temperature perturbation eigenfunction at optical depth $\tau=2/3$ which can be calculated using an adiabatic oscillation code. The product, $K_\alpha \delta F_{n\ell}(R_*)$, which contains the dependence of the observed flux variation on the structure of the secondary star, will be calculated for several different stars in §5.

We next calculate the change in the projected surface area of the distorted star associated with different modes. The change in the area also causes the observed flux to vary. The distortion of the surface due to one mode can be found from the displacement eigenfunction, which can be written as

$$\boldsymbol{\xi}_{n\ell m}(r,\theta,\phi) = \left\{ \xi_{n\ell}^r(r), \xi_{n\ell}^h(r)\frac{\partial}{\partial\theta}, \xi_{n\ell}^h(r)\frac{\partial}{\sin\theta\partial\phi} \right\} Y_{\ell m}(\theta,\phi), \tag{34}$$

where $\xi_{n\ell}^r$ and $\xi_{n\ell}^h$ are the radial and transverse displacement eigenfunctions, respectively. These eigenfunctions are normalized so that the energy in the mode is unity, i.e.,

$$\omega_\alpha^2 \int d^3x\, \rho\, \boldsymbol{\xi}_{n\ell m} \cdot \boldsymbol{\xi}_{n\ell m} = 1. \tag{35}$$

The fractional change in the projected area of the star associated with a mode, $\delta S_{n\ell m}/\pi R_*^2$, can be expressed in terms of the surface displacement. The fulx variation arising from this change in projected area can be shown to be identical to the expression for $\delta F_{n2m}^{obs}/F^{obs}$ of equations (31) and (32), provided that we replace $a_{LD}[\delta F_{n2}(R_*)/F]$ in these equations by

$$\left[ \frac{8\xi_{n2}^r(R_*) - 6\xi_{n2}^h(R_*)}{R_*} \right].$$

This allows us to define an effective flux variation eigenfunction $\delta F_{n2}^{eff}$ as

$$\frac{\delta F_{n2}^{eff}}{F} \equiv \frac{\delta F_{n2}}{F} + \frac{2}{a_{LD}} \left[ \frac{4\xi_{n2}^r(R_*) - 3\xi_{n2}^h(R_*)}{R_*} \right] \tag{36}$$

The total observed flux variation associated with modes is obtained from equations (31) and (32) by replacing $\delta F_{n2}/F$ in these equations with the effective flux eigenfunction as given by the above equation.

§**3.2 Surface velocity amplitude associated with modes**

The calculation of the surface velocity proceeds in a similar manner to the flux variation calculation. The line-of-sight surface velocity, integrated over the stellar hemisphere facing the observer, is given by

$$V_{n\ell m}^{obs} = \frac{\omega_\alpha A_{n\ell m}}{\pi} \int d\Omega\, |\mathbf{n}_s \cdot \mathbf{n}_{obs}|\, \mathbf{n}_{obs} \cdot \boldsymbol{\xi}_{n\ell m}(R_*), \tag{37}$$

where the first factor in the above integrand weighs the velocity by the projected area.

Carrying out the integral given in equation (37) for $\ell = 2$, we obtain the following expression for the observed velocity amplitude for an $m=0$ mode

$$V_{n20}^{obs} = f_{cons} \sqrt{\frac{1}{45\pi}} \omega_{n2} K_{n2} \left[ \xi_{n2}^r(R_*) + 3\xi_{n2}^h(R_*) \right] \tilde{A}_{n20}[3\cos^2 i - 1] \tag{38}$$

and for the sum of the $m = \pm 2$ modes

$$V_{n22}^{obs} = f_{cons}\sqrt{\frac{2}{15\pi}}\omega_{n2}K_{n2}\left[\xi_{n2}^r(R_*) + 3\xi_{n2}^h(R_*)\right]\sin^2 i\left[Re(\tilde{A}_{n22})\cos(2\phi_o - 2\Omega_*t) - Im(\tilde{A}_{n22})\sin(2\phi_o - 2\Omega_*t)\right]. \tag{39}$$

These expressions have the same functional dependence as the flux variation derived in the previous subsection.

### §3.3 Perturbation to pulsar orbit due to pulsation

The contribution to the gravitational potential at the position of the primary star due to one mode of pulsation of the secondary can be easily calculated using equation (1) and is given below

$$\delta\Psi_\alpha = -\frac{4\pi G Q_\alpha Y_{\ell m}(\pi/2, \phi_{orb}(t))A_\alpha(t)}{(2\ell + 1)R^{\ell+1}}. \tag{40}$$

The equation for the perturbation to the primary's position, $\mathbf{R}_1$, due to this potential, to first order in $\mathbf{R}_1$, is as follows

$$\frac{d^2\mathbf{R}_1}{dt^2} = -\frac{\Omega_o^2}{(R_0/a)^3}\left[\mathbf{R}_1 - 3\hat{\mathbf{R}}_0(\hat{\mathbf{R}}_0 \cdot \mathbf{R}_1)\right] - \frac{M_t}{M_*}\boldsymbol{\nabla}\delta\Psi_\alpha, \tag{41}$$

where $\mathbf{R}_0$ is the primary's unperturbed position with respect to the center of mass, $\hat{\mathbf{R}}_0 = \mathbf{R}_0/|\mathbf{R}_0|$, and $M_t$ is the total mass in the binary system. The first term in the above equation arises due to the change in the gravitational force on the primary star because of the perturbation to its position. In general the solution of equation (41) must be obtained numerically. However, for modes of small $b_\alpha$ the mode amplitude $A_\alpha$ is a sinusoidal function, and if $r_\alpha \gg 1$ (highly eccentric orbit), the first term on the right side of the above equation can be ignored, and equation (41) can be solved to yield the ratio of the perturbed to the unperturbed velocity of the primary. The result below is for quadrupole modes

$$\frac{|\omega_\alpha \mathbf{R}_1|}{a\Omega_o} \approx \frac{16\pi^2}{25}\left(\frac{\Omega_o\omega_\alpha Q_\alpha^2 M_p}{M_t M_*}\right)\left[\frac{\tilde{A}_\alpha(t) Y_{\ell m}(\pi/2, \phi_{orb}(t))}{a^2(R/a)^4}\right]. \tag{42}$$

Using results from §2.2, we find that the fractional perturbation to the primary's orbital velocity is typically very small.

It is useful to express the perturbation to the orbit as a delay in the pulse arrival time, $\delta t_\alpha$, when the primary is a pulsar. This delay is a periodic function and is given below

$$\delta t_\alpha = \sin i\ (\mathbf{R}_1 \cdot \mathbf{n}_o)/c,$$

where $\mathbf{n}_o$, a unit vector, is the projection of the observer's direction on the orbital plane.

## §4 Determination of orbital inclination angle

We describe a technique for determining the inclination angle of an eccentric orbit by fitting the observed light curve near periastron to theoretical light curves. The main physical idea behind this method is simple. In the static tide limit, when periods of low order modes are small compared to the periastron passage time, the star adjusts its shape to the instantaneous tidal force and thus the tidal bulge points in the direction of the other star irrespective of its rotation rate. Therefore, as the secondary star makes its passage through periastron, the tidal bulge continues to point toward the primary star, and the time dependence of the tidally distorted surface of the star projected onto the plane of the sky gives rise to the observed light curve. Since the observed part of the surface of the star depends on the inclination angle, the shape of the light curve also depends on the angle $i$.

For an orbital inclination angle, $i$, and a longitude of periastron, $\omega$, the observed flux variation, $\delta F_{obs}$, is the sum of the contribution from $m = \pm 2$ and $m = 0$ modes and can be written in the following form (see eqs. [31] and [32])

$$\delta F_{obs}(t) \propto \sin^2 i \left[ \sin(2\omega - 2\Omega_* t)\Psi_2(t) - \cos(2\omega - 2\Omega_* t)\Psi_1(t) \right] + (3\cos^2 i - 1)\Psi_0(t), \quad (43)$$

where $\Psi_1$ and $\Psi_2$ are the real and imaginary parts of the mode amplitude summed over all $m = 2$ modes (weighted by $K_{n2}\delta F_{n2}^{eff}$), as seen in the rotating frame of the star, and $\Psi_0$ is the sum over all $m = 0$ modes. In the limit of static tide, the tidal bulge always points towards the other star; therefore, the observed flux variation is independent of the stellar rotation rate $\Omega_*$ (see the derivation below). We see from equation (43) that the *shape* of the observed light curve in general depends on the orbital inclination angle because of the different relative contributions from $m = 0$ and $m = \pm 2$ modes. For the particular case of circular orbits, however, the amplitudes of $m=0$ modes are time independent so that the *shape* of light curve is independent of $i$. Of course, the magnitude of the observed flux variation still depends on $i$, but because of the uncertainty in stellar mass and structure this is of limited use in determining $i$.

If the dimensionless mode frequency, $b_\alpha$, for the first few $p$ and $g$-modes is greater than about 5, then the time derivative terms in equation (2) can be neglected and so the mode amplitudes are equal to $f_\alpha/\omega_\alpha^2$. In this static tide limit the function $\Psi_0(t)$ is proportional to the reduced forcing function $\tilde{f}_{20}$, and $\Psi_1(t)$ & $\Psi_2(t)$ are proportional to the real and imaginary parts of $\tilde{f}_{22}$ (it is easy to show that the tidal bulge points in the direction of the other star in this case). Thus using equations (5) and (43) we obtain the following explicit expression for the light curve

$$\delta F_{obs}(t) = -F_c \left[ \frac{1 - 3\sin^2 i \sin^2(\phi_{orb}(t) - \omega)}{(R(t)/a)^3} \right], \quad (44)$$

where $F_c$ is a constant factor that depends on the stellar mass and structure and is independent of the orbital parameters. $F_c$ can be calculated either by determining the normal mode eigenfunctions and the associated flux variation at the surface or using von Zeipel's theorem (cf. Kopal 1959), according to which the emergent radiation flux, in the static tidal case, is proportional to the local gravity. We note that the rotation rate of the star has dropped out from the above equation for the observed flux variation as was claimed earlier. Thus a comparison of the *shape* of the observed light curve with the function given in equation (44) provides a robust method, which does not depend on the unknown mass, structure and rotation rate of the star, for determining the orbital inclination angle.

Light curves, corresponding to several different inclination angles, for a $10M_\odot$ main sequence star in an orbit of $e=0.4$ and period of 100 days are shown figure 2. The value of $\phi_0$ is chosen to be $0°$. We emphasize that the magnitude of the flux variation in these graphs depends on the mass and structure of the star as well as the inclination angle. However the shape, which is what we suggest should be used to compare with the observed light curve to determine the inclination angle, does not. We note that for small $i$, the contribution to the observed light curve is dominated by $m=0$ modes. As $i$ increases, the contribution from the $m=\pm 2$ modes increases, resulting in a significant change in the shape of the light curve. The shape of the light curve is most sensitive at intermediate values of $i$ ($20 \lesssim i \lesssim 70$); however, there are quite noticeable variations for all $i$.

The order of magnitude of the flux variation can be estimated crudely from the change in the Lagrangian temperature at the point in the star's atmosphere where the optical depth is 2/3. Since the tide is very nearly static, we can integrate the hydrostatic equilibrium equation from the surface over a fixed column of gas corresponding to an optical depth of 2/3, with and without the tidal force, to obtain the fractional change in the Lagrangian pressure and find it to be $\delta g/g + F_t/g = 2\delta R_*/R_* + 2(M_p/M_*)(R_*/d)^3$, where $g$ is the gravitational acceleration at the stellar surface, $F_t$ is the tidal acceleration, $d$ is the distance between the stars, and $\delta R_*$ is the change in star's radius, which is approximately equal to $(M_p/M_*)(R_*/d)^3 R_*$. Thus the fractional change in the luminosity of the SMC star, for which $M_p/M_* \sim 1/7$ and $R_*/d \sim 0.2$ at periastron, is estimated to be about 0.5% or 5 m-mag. Our numerical value is a factor of about 2 larger (see §5.1).

The so called ellipsoidal variable stars are believed to undergo flux variations due to tidal distortion which are correlated with the orbital phase (c.f. Kopal 1959, Avni & Bahcall 1975, Morris 1985). It is well known that the magnitude of the observed light variation of the ellipsoidal variable stars depends on the orbital inclination angle. However, it is not appreciated that the *shape* of their light curve also depends on the orbital inclination angle if the orbit is non-circular. Finally, we note that our proposed method for determining orbital inclination angle is expected to work best in those systems where one of the stars is compact so that the interpretation of light curve is not complicated by eclipses.

## §5 Applications

In this section, we apply our formalism to two radio binary pulsar systems, PSR J0045-7319 and PSR B1259-63, and discuss the possibility of direct observation of the tidal excitation of modes in these systems. We also present results of numerical calculations for general polytropic stars in order to further examine the dependence of tidal excitation on stellar structure (§5.2).

### §5.1 Applications to PSR J0045-7319 and PSR B1259-63

Recently, Kaspi et al. (1994) have reported the discovery of a radio pulsar (PSR J0045-7319) in the Small Magellanic Cloud. Measurements of the Doppler time delay have indicated that the pulsar is in a highly eccentric ($e = 0.80798$) orbit with a period of 51.169 days. Kaspi et al. (1994) have optically identified the companion star as a 16th magnitude B-star with mass of $\approx 10 M_\odot$. In our calculations, we also assume that the mass of the pulsar is $1.4 M_\odot$.

Using a B-star model kindly provided to us by the Yale group, we have computed the eigenfrequencies and the overlap integrals ($Q_{nl}$) for quadrupole oscillations. We find that the $f$-mode frequency for this system corresponds to a $b_\alpha$ value of approximately 21.7, and the $g_5$ mode has a $b_\alpha$ value of 3.96. The frequencies of $m = \pm 2$ modes are shifted by approximately $\pm 1 \mu$Hz, if the B-star is rotating synchronously at periastron, which corresponds to a shift in the value of $r_\alpha$ of about $\pm 4.4$. The radiative damping time depends on the mode order: for $g_2$ and $g_3$-modes we estimate it to be about $10^3$ years, and for $g_{10}$ mode to be about one year. The turbulent damping time for these modes is a factor of a few larger. The radiative damping time for the $p_2$ mode is about a year. Thus for none of these modes do we expect the mode energy to have any dependence on damping.

For the $10 M_\odot$ ZAMS stellar model the product $K_{n2} \, \delta F_{n2}(R_*)/F$ is very nearly independent of $n$ for $g$-modes of order 4 to 10, beyond which this product starts to decrease. The probability that one of the approximately 25 low to intermediate order g-modes (counting m=0, and $\pm 2$ as three distinct modes due to their different frequencies because of rotational splitting) has a value of $r_\alpha$ that is close to resonance ($|\delta r_\alpha| \lesssim 0.05$) is very high. In fact, for the stellar model we use, the $g_7$ mode has $\delta r_\alpha$ of 0.02, which is the smallest value for the first 10 harmonic $g$ modes. The smallest physically meaningful value of $|\delta r_\alpha|$, resulting from the orbital evolution considerations (see §2.4), is estimated using equation (26) to be 0.003. Thus, $\delta r_\alpha = 0.02$ for the $g_7$ mode is physically acceptable. The $g_7$ mode, which has the largest amplitude, has a period of 1.07 days (corresponding to $b_\alpha = 2.96$). The stellar flux variation associated with this mode is 2.3 m-mag, and the surface velocity amplitude is 70 m s$^{-1}$ (calculated using eqs. [38] and [39]). For comparison, the flux variation associated with the $f$-mode, away from periastron, is about 10 $\mu$-mag.

The perturbation to the pulsar orbital velocity due to the time dependent quadrupole moment of the $g_7$ mode has been calculated by integrating equation (41) and is found to be $10^{-3}$ cm s$^{-1}$. The density fluctuation in the convection zone of the star also causes a random perturbation to the pulsar's orbital velocity that is estimated to be about $10^{-6}$ cm s$^{-1}$ (see appendix).

It should be stressed that the amplitudes of these tidally excited modes (shown in Fig.1) have perfect phase coherence from one orbit to another which should be helpful in observing these small amplitude oscillations. On the other hand, since the mode with the largest observed amplitude is one with a frequency closest to resonance, it would be very difficult to make a precise identification of the order ($n$) of the observed mode, rendering its potential use in determining the properties of the star rather limited.

The light curve for PSR J0045-7319 near periastron, dominated by $f$ and low order $p$-modes, is shown in figure 3 for several different orbital inclination angles. The magnitude of the observed flux variation is about 10 m-mag and is calculated assuming that the secondary star has a mass of 10 $M_\odot$. The uncertainty in our estimate of the flux variation is about 20%, reflecting an uncertainty in the mass of the B-star of $2M_\odot$. The *shape* of the light curves in figure 3, however, is independent of the mass of the star, and can be used to determine the inclination angle of the orbit.

We have calculated the periastron advance of the PSR J0045-7319 system, due to dynamical tides, by integrating equation (36) and summing the result over all modes. The advance is found to be 0.7 arcsec per orbit. This result is in excellent agreement with the classical apsidal motion formula (Claret & Giménez, 1993); our result is about 1% larger compared to the classical formula. Thus the static tide approximation for calculating apsidal motion is an excellent approximation for this system.

The recent accurate observations of PSR B1259-63 (Johnston et al. 1994) indicate that its eccentricity and orbital period are approximately 0.8698 and 1236.79 days, respectively. The companion is a Be-star with a mass of $\approx 10 M_\odot$. Thus the periastron distance is $\approx 20 R_*$, and the $b_\alpha$ values for the modes of the Be-star are about a factor of 13 larger than those for the SMC B-star. Therefore, the amplitude near periastron, which falls off as the distance cubed, is expected to be a factor of about 70 smaller than that of the SMC system, and the oscillation amplitudes of even the $g$ modes, because of the large $b_\alpha$ values, are too small to have any observational consequences.

### §5.2 Application to Polytropic Stars

In this section, we apply our discussion of mode amplitude to a general polytropic star. In particular, we examine how the flux variation and surface velocity depend on the structure of the star. Our calculations include the $f$-mode as well as low order $p$-modes and $g$-modes for polytropic stars with index ranging from $n_{pi} = 1.5$ to 3.

For a polytropic star, the eigenfrequencies and overlap integrals can be expressed in nondimensionalized forms which depend only on the index $n_{pi}$. We have calculated these quantities without the Cowling approximation, and our results are in excellent agreement with values quoted in Robe (1968) and Lee and Ostriker (1986).

The dependence of the mode energy on the stellar structure is contained in the parameter $K_\alpha^2$. We calculate this numerically as a function of polytropic index, $n_{pi}$, for the quadrupole $f$-mode and low order $p$-modes and $g$-modes. The results are plotted in figure 4 in the units of $GM_*^2/R_*$. Mode energy is obtained by multiplying $K_{n2}^2$, $|\tilde{A}_{n2m}|^2$ (calculated using eq. [10]), and $f_{cons}^2$. While the $f$-mode and the $p$-modes have comparable values of $K_\alpha^2$, the $g$-modes are few orders of magnitude smaller. This is a result of two separate effects. First, the overlap integral is largest for the $f$-mode and decreases with increasing mode order $n$. Second, the eigenfrequencies for the $p$-modes are much larger than those for the $g$-modes. Thus, even though $p$-modes and $g$-modes could have comparable overlap integrals, $K_\alpha$ still differs significantly. Figure 4 shows that $K_\alpha^2$ increases with $n_{pi}$ for all modes. This might suggest that higher $n_{pi}$ polytropes are more susceptible to tidal excitations. However, it is important to keep in mind that the mode energy is also proportional to $|\tilde{A}_\alpha|^2$, which has a sensitive dependence on the mode frequency through $b_\alpha$; $|\tilde{A}_\alpha|^2$ decreases exponentially with $b_\alpha$. Since for a fixed mode, the mode frequency (and hence $b_\alpha$) tends to increase with $n_{pi}$, $\tilde{A}_\alpha$ tends to decrease with $n_{pi}$. Therefore, depending on the orbital parameters, the actual mode energy might in fact decrease with $n_{pi}$.

The fractional flux variation from the star depends on the dimensionless parameter $K_\alpha \delta F_{n\ell}(R_*)/F$, which is approximately equal to $4K_\alpha \Delta T(\tau = 2/3)/T$, where $T(\tau = 2/3)$ is the Lagrangian temperature variation at optical depth of 2/3. We calculate $\Delta T$ numerically using an adiabatic oscillation code. This is then used to compute $K_\alpha \delta F_{n\ell}(R_*)/F$, and the results are shown in figure 5 for several different quadrupole modes as a function of $n_{pi}$ for a fixed stellar mass of 1.0 $M_\odot$. The $p$-modes correspond to acoustical compression or expansion, while the $g$-modes are effectively incompressible. Hence, $p$-modes have correspondingly larger values of $\Delta T/T$. For polytropic stars, the dimensionless parameter $K_\alpha \delta F_{n\ell}(R_*)/F$ is found to be almost independent of the mass of the star.

The surface velocity amplitude depends on the parameters $K_\alpha \omega_\alpha \xi_\alpha^r$ and $K_\alpha \omega_\alpha \xi_\alpha^h$ (see eqs. [38] and [39]). These parameters are shown in figure 6 in the units of $\sqrt{GM_*/R_*}$. The $p$-modes have predominantly radial displacement at the surface, while $g$-modes have predominantly transverse displacement.

The shift of the mode frequencies due to rotation can be calculated using the dimen-sionless number $J_{n2}$ (see eq.[16]) which is shown in figure 7, for several different modes, as a function of $n_{pi}$. For $p$-modes, $J_{n2}$ approaches 0 as $n$ increases since the transverse displacement becomes negligibly small compared with the radial displacement. For $g$-modes, it is the radial displacement that becomes negligible as $n$ increases; hence $J_{n2}$ approaches $1/\ell(\ell+1)$ or 0.1667 for quadrupole modes (see, e.g., Unno et al. 1989). Therefore, $J_{n2}$ is bounded above by the $f$-mode and is always less than 0.5.

In order to obtain the actual flux variation or surface velocity, one needs to multiply the relevant products (displayed in Figs. 5 and 6) by the reduced mode amplitude $\tilde{A}_\alpha$, which depends on the orbital and modal parameters through $r_\alpha$ and $e$. As an example, consider a binary system with $e = 0.7$, consisting of a 1.0 $M_\odot$ compact primary, and a 5.0 $M_\odot$ secondary star of polytropic index 2. To find the flux variations associated with the $p_1$ and $g_3$-modes of the secondary, we read off from figure 5 that the values of $K_\alpha \delta F_{n\ell}/F$ are approximately 100 and 0.1 respectively. Assume that the orbital period is such that the $g_3$ mode has $r_\alpha$=23.8 ($b_\alpha$=3), and the $p_1$ mode has $r_\alpha$=188.5 ($b_\alpha$=23.8). Using the procedures outlined in §2, we calculate $\tilde{A}_\alpha$ for the $g_3$ ($p_1$) mode with $m$=0 and find the peak value of the reduced amplitude at periastron to be 0.02 ($3.3 \times 10^{-4}$) and an oscillation amplitude of 0.01 ($1.3 \times 10^{-7}$) away from periastron. The product of $\tilde{A}_\alpha$, $K_\alpha \delta F_{n\ell}/F$, and $f_{cons}$ (=2.1) yields the flux variations for these modes to be 20 (33) milli-magnitudes at periastron and 10 (0.013) milli-magnitudes for the oscillations away from periastron.

For concreteness, we have calculated in figure 8 the fractional luminosity variations, $\delta F_{n22}^{obs}/F$, for several low-order $g$-modes in a $n_{pi} = 3$ polytrope as a function of orbital period. To prevent close resonances from confusing the general trend, we have modified the mode frequencies slightly such that $\delta r_\alpha = 0.1$ for all the modes considered. The mass of the secondary polytropic star is taken to be 5 $M_\odot$, and the mass of the primary star is 2 $M_\odot$. To illustrate the dependence on eccentricity, $e$ has been chosen to be 0.8 and 0.4 (Fig. 8a and Fig. 8b, respectively). Figure 8 shows that $\delta F_{n22}^{obs}/F$ falls off rapidly with the orbital period and that it is significantly larger for more eccentric orbits. For larger orbital periods, variations due to higher overtone $g$-modes dominate because of their smaller $b_\alpha$ values.

## §6. Summary and discussion

We have calculated the energy and amplitude of the tidally excited modes of a star in a binary system of arbitrary eccentricity. The effects of damping, resonance, orbital inclination, and rotation of the star on the observed mode amplitude have been investigated in detail. The important parameters for this problem are: $b_\alpha$, the ratio of the periastron passage time to the mode period; $\delta r_\alpha$, the ratio of the difference between mode frequency and the nearest resonance frequency (resonance frequencies are integer multiples of the orbital frequency) to the orbital frequency; and $d_\alpha$, the ratio of the orbital period to the mode damping time.

Modes of large $b_\alpha$ couple poorly to the tidal force. The oscillation amplitude decreases with increasing $b_\alpha$ as approximately $\exp(-1.3 b_\alpha e^{-0.25})$, where $e$ is the orbital eccentricity. The $p$, $f$, and low order $g$-modes of main sequence stars, for most binary systems, have values of $b_\alpha$ that are much greater than 1; thus, their oscillation amplitudes are exponentially suppressed. Since the overlap integrals decrease with increasing mode order ($n$), the moderate order quadrupole $g$-modes with $b_\alpha \sim 1$ and the smallest value of $\delta r_\alpha$, i.e., the modes most nearly resonant, have the largest oscillation amplitudes.

The mode amplitude is independent of damping so long as $d_\alpha$ is less than $\delta r_\alpha$. This condition is expected to be satisfied for the low to moderate order $g$ modes of main sequence stars which have damping times of at least a few years. In the limit of $d_\alpha \ll \delta r_\alpha$ the mode amplitude is proportional to $1/\delta r_\alpha$, as expected of an oscillator driven off-resonance. Due to orbital evolution, a mode which is in perfect resonance with the orbit, i.e. $\delta r_\alpha = 0$, moves off resonance with time. This places a severe upper limit to the amplitudes of modes near resonance (§2.4).

The mode amplitude and phase of tidally excited modes is the same from one orbit to another. This property should be useful in observational searches for tidally excited modes as it distinguishes them from intrinsically excited modes. Rotation of the star (in the limit of slow rotation rate) causes the frequencies of $m \neq 0$ modes to be shifted, but otherwise does not change the properties or time dependence of the mode amplitude.

We have applied this tidal excitation theory to the recently discovered pulsar in the SMC, PSR J0045-7319, which is in a binary system of eccentricity 0.81 and orbital period of 51 days. The companion star, believed to be a $10 M_\odot$ B-star, has a periastron distance of about 5 stellar radii. Thus this system is a good candidate for studying the tidal excitation of modes. Using a $10 M_\odot$ main sequence stellar model, we find that the mode with the largest oscillation amplitude is the $g_7$ mode, which has $b_\alpha = 2.96$ and $\delta r_\alpha = 0.02$. We find that the flux variation associated with this mode is 2.3 milli-magnitude, and the surface velocity amplitude is 70 meter s$^{-1}$. The perturbation to the orbital velocity of the pulsar, as a result of the time dependent quadrupole moment of the B-star due to this mode, is about $10^{-3}$ cm s$^{-1}$ at periastron. These amplitudes are independent of mode dissipation time because the value of $d_\alpha$ is much less than $\delta r_\alpha$ (the radiative and turbulent dissipation time scales are about 100 years).

It should be pointed out that the small value of $\delta r_\alpha$ for the $g_7$ mode is a coincidence. A change of 0.2% in the mode frequency changes $\delta r_\alpha$ by 0.1. Thus $\delta r_\alpha$ should be regarded as a random variable. Therefore, given 25 modes that are similar in terms of their coupling to the tidal force, we expect one of them to have $|\delta r_\alpha|$ less than 0.02. In fact, modes $g_4$–$g_{10}$ of the Yale star model have very similar mode amplitudes for a given $\delta r_\alpha$. Thus it is not surprising to find one of them, counting $m=0, \pm 2$ as three different modes because rotation lifts the $m$-degeneracy of the frequencies, with $\delta r_\alpha$ of 0.02. Orbital evolution places a lower limit on $|\delta r_\alpha|$, which we estimate to be 0.003.

The tidal distortion of the star near periastron is dominated by low order quadrupole $p$ and $g$-modes and the $f$-mode, for which $b_\alpha$ is much greater than one, and therefore the light curve near periastron follows the time dependence of the tidal force closely. Since the observed light curve depends on the projection of the tidally distorted star onto the plane of the sky, which in turn depends on the inclination of the orbit, the shape of the light curve near periastron allows the determination of the inclination angle of the orbit for eccentric orbits. In particular, this technique should be applicable to the SMC pulsar, PSR J0045-7319, for which we estimate the flux variation near periastron to be $\approx$ 10 m-mag, and the surface velocity to be about 30 m s$^{-1}$.

For the SMC pulsar, we further determine that the periastron advance due to the tidal distortion of the B-star, treated dynamically, is about 0.7 arcsec per orbit. This is in excellent agreement ($\approx$ 1.0 %) with the static tidal calculation of the classical apsidal motion formula. The precession due to general relativity is estimated to be 2.0 arcsec per orbit and that due to the quadrupole moment due to rotational distortion (for synchronous rotation) is about 2.2 arcsec/orbit. Thus the total apsidal motion compares reasonably well with the current observational value of about 5 arcsec per orbit (Kaspi et al. 1994).

For the only other known binary radio pulsar with a main sequence companion, PSR B1259-63, the flux variation near periastron is approximately 70 times smaller than that of the SMC system, and the oscillation amplitude of all $g$ and $p$-modes is negligibly small.

We have also calculated the stellar structure dependence of the mode energy, flux variation, and surface velocity amplitude by considering polytropic stars of index $n_{pi}$ ranging from 1.5 to 3. These quantities can be written as a product of the two terms: a term that only depends on the reduced mode amplitude ($\tilde{A}_\alpha$) and a term that depends on the stellar structure ($K_\alpha, \delta F_\alpha/F$, etc.). The stellar structure dependent term generally increases with $n_{pi}$. However, this does not necessarily mean that higher $n_{pi}$ stars give rise to larger tidal effects. The mode frequency (and thus $b_\alpha$) increases with $n_{pi}$, and consequently $\tilde{A}_\alpha$ decreases with $n_{pi}$. Therefore, depending on the orbital parameters, tidal effects might in fact be larger for lower $n_{pi}$ stars.

**Acknowledgment:** We are indebted to Paul Schechter for many excellent suggestions and encouragements and to Pierre Demarque, Sydney Barnes & David Guenther for providing stellar models. We thank Vicky Kaspi, Alan Levine, Jerry Ostriker, Saul Rappaport and Alar Toomre for helpful discussions. We also thank the referee Andreas Reisenegger for several valuable suggestions on the original draft. This work was supported in part by a NASA grant NAGW-3936.

# Appendix

In this appendix, we estimate the perturbation to the primary star's orbital velocity due to density fluctuations in the convection zone of the secondary star.

We first calculate the quadrupole moment of the star due to fluctuating eddies in one scale height thick layer ($H$) of fluid at a radial distance $r$ from the center. Let the mean mass of this layer be $M_H$ and the magnitude of the fluctuation be $\delta M_H$. Since the fractional density fluctuation, $\delta \rho/\rho$, associated with turbulent flow is $\mathcal{M}_H^2$, where $\mathcal{M}_H$ is the Mach number of the turbulence, therefore the fractional mass fluctuation of this layer is $\delta M_H/M_H \sim H \mathcal{M}_H^2/r$. Let us assume that the luminosity carried by convection is $f L_*$, where $L_*$ is the luminosity of the star. According to mixing length theory $f L_* \sim 4\pi \rho r^2 C^3 \mathcal{M}_H^3 \sim M_H C^3 \mathcal{M}_H^3/H$, where $C \sim \sqrt{gH}$ is the sound speed and $g$ is gravitational acceleration. Thus the RMS fluctuation of the quadrupole moment $\delta Q_H \sim r^2 \delta M_H \sim r M_H^{1/3}(fL_* H)^{2/3}/g$. Since $\delta Q_H \propto (H^2 M_H)^{1/3} \propto (\rho H^3)$, the contribution to the quadrupole moment is dominated by eddies at the bottom of the convection zone. So from here on we will take $H$ to be the scale height at the bottom of the convection zone. The perturbation to the gravitational potential due to the quadrupole moment $\delta Q_H$ is

$$\delta \Psi \approx -\frac{4\pi G \delta Q_H}{5 R^3}.$$

We can determine the perturbation to the primary's orbital velocity by solving equation (41) with $\delta \Psi$ given in the above equation. If the convective time scale is much smaller than the orbital period, so that orbital modes are not resonantly excited by convection, we can discard the first term in equation (41), and the solution of the resultant equation is given below

$$\frac{1}{\Omega_o a}\left|\frac{d\mathbf{R}_1}{dt}\right| \sim f f_r^3 (\Omega_o t_H)^2 \left(\frac{R_*}{R}\right)^2 \left[\frac{a}{R}\right]^2 \left(\frac{1}{\Omega_o t_{kh}}\right),$$

where $f_r = r/R_*$, $t_H = H/V_H$ is the convective time, $V_H$ is the convective velocity, and $t_{kh} = GM_*^2/(L_* R_*)$ is the Kelvin-Helmholtz time scale of the secondary star. The dimensionless factors $f$ and $f_r$ are roughly of order 1. Therefore the random fluctuation to the primary's orbital velocity, due to a $10 M_\odot$ main sequence secondary star in an orbit with $a/R_* \sim 25$, is a factor of $10^{13}$ smaller than the mean orbital velocity.

**Figure Captions**

FIG. 1.—The reduced mode amplitude $\tilde{A}_\alpha$ as a function of time in one orbit, shown for several values of $b_\alpha$ for $m=0$. Here we take $e = 0.8$, $\delta r_\alpha = 0.1$, and $d_\alpha = 0.01$. $t = 0$ corresponds to periastron. Note that the oscillation amplitude falls off exponentially with $b_\alpha$, whereas the peak at periastron falls off more slowly. For $b_\alpha = 2$, the oscillation amplitude is comparable to the periastron amplitude. For $b_\alpha = 10$, the oscillation amplitude is so small that the oscillations can hardly be seen in Fig. 1c. In Fig. 1d, a portion of Fig. 1c near apstron is expanded in scale to show the oscillations more clearly. These tidally excited oscillations have phase coherence with the orbital motion; as such, they should be readily distinguished from intrinsic oscillations of the star. The $m = \pm 2$ modes have similar behaviors but tend to have larger oscillation amplitudes for a fixed $b_\alpha$.

FIG. 2.—Light curve near periastron for a 10 $M_\odot$ star in an orbit with $e = 0.4$ and $T = 100$ days for several values of the orbital inclination angles $i$; the unit of flux variation is arbitrary, and $t = 0$ corresponds to periastron. The value of $\phi_0$ is chosen to be $0^o$. The shape of the light curve varies as a function of $i$ due to the relative contributions from the $m = 0$ and $m = \pm 2$ modes, which project differently onto the plane of the sky. While the magnitude of these curves depends on the mass and detailed structure of the star, the shape does not. Hence it can be used to determine $i$ even when the stellar structure of the star is poorly known.

FIG. 3.—Light curve near periastron for the B-star in the SMC pulsar system (PSR J0045-7319) for several orbital inclination angles. $\phi_0$ is taken to be $25.24^o$, corresponding to a longitude of periastron of $\omega = 115.24^o$, as quoted in Kaspi et al. (1994). The shape of these curves can be used to determine the inclination angle for this system.

FIG. 4.—$K_\alpha^2$ in the units of $GM_*^2/R_*$ as a function of polytropic index, $n_{pi}$, for the quadrupole $f$-mode and low order $p$ and $g$-modes. The parameter $K_\alpha^2$ contains the dependence of the mode energy on stellar structure. Note that $K_\alpha^2$ for $g$-modes is a few orders of magnitude smaller than for $p$-modes even though their overlap integrals are comparable. This is because the eigenfrequencies for the $g$-modes are much smaller than those for the $p$-modes.

FIG. 5.—$|K_\alpha \delta F_{n\ell}/F|$ as a function of polytropic index, $n_{pi}$, with stellar mass fixed at $1 M_\odot$. This dimensionless parameter contains the dependence of the fractional luminosity change on the stellar structure. Since $p$-modes are compressible whereas $g$-modes are effectively incompressible, the $p$-modes have larger values of the Lagrangian temperature perturbation, $\Delta T/T$, leading to larger values of $K_\alpha \delta F_{n\ell}/F$. Furthermore, this parameter varies monotonically with $n_{pi}$ for $p$-modes, but has a minimum for $g$-modes corresponding to a node at the surface for $\Delta T$.

FIG. 6.—$|K_\alpha \omega_\alpha \xi_\alpha^r|$ and $|K_\alpha \omega_\alpha \xi_\alpha^h|$ in the units of $\sqrt{GM_*/R_*}$ as a function of polytropic index, $n_{pi}$. These two parameters contain the dependence of the surface velocity amplitude on the stellar structure. The $p$-modes have predominantly radial displacement, while $g$-modes have predominantly transverse displacement.

FIG. 7.—$J_{n2}$ as a function of polytropic index, $n_{pi}$. Rotational splitting of the eigenfrequencies is, to first order in $\Omega_*$, proportional to $J_{n\ell}$ (see eq. [16]). For $p$-modes, $J_{n2}$ approaches 0 as the order $n$ increases. For $g$-modes, $J_{n2}$ approaches 1/6 as $n$ increases.

FIG. 8.—Fractional luminosity variation $\delta F_{n22}^{obs}/F$ for several low-order $g$-modes as a function of orbital period $T_{orb}$. The secondary star is taken to be a $n_{pi} = 3$ polytrope with $M_* = 5 M_\odot$, and the primary star has $M_p = 2 M_\odot$. For all the modes considered, $\delta r_\alpha$ is fixed artificially at 0.1, to avoid confusion due to close resonances, and $d_\alpha = 0.001$. The eccentricity of the orbit is taken to be 0.8 and 0.4 (Fig. 8a and Fig 8b, respectively). The lower limits for $T_{orb}$ of approximately 18 days (for $e = 0.8$) and 4 days (for $e = 0.4$) correspond to the onset of the overflow of the secondary's Roche lobe. For larger $T_{orb}$, the luminosity variation due to the higher overtone $g$ modes dominates because of their smaller $b_\alpha$ values. Morever, the variations are significantly larger for the more eccentric orbit.